\documentclass[letterpaper,twocolumn,10pt]{article}
\usepackage{usenix2019_v3}

\usepackage{tikz}
\usepackage{amsmath}
\usepackage{listings}

\usepackage{filecontents}
\usepackage{algorithm}
\usepackage{algorithmic}
\usepackage{tabularx,booktabs}
\usepackage{caption}
\usepackage{subcaption}

\newcommand{\system}{Shabari}

\begin{document}

\date{}

\title{\Large \bf Shabari: Delayed Decision-Making for Faster and Efficient Serverless Functions}

\author{
{\rm Prasoon Sinha}\\
The University of Texas at Austin
\and
{\rm Kostis Kaffes}\\
Columbia University
\and
{\rm Neeraja J. Yadwadkar}\\
The University of Texas at Austin and VMWare Research
} 

\maketitle

\begin{abstract}
Serverless computing relieves developers from the burden of resource management, 
thus providing ease-of-use to the users and the opportunity to optimize resource utilization for the providers. 
However, today's serverless systems lack performance guarantees for function invocations, thus limiting support for performance-critical applications: we observed severe performance variability (up to 6x).
Providers lack visibility into user functions and hence find it challenging to right-size them: 
we observed heavy resource underutilization (up to 80\%). 
To understand the causes behind the performance variability and underutilization, we conducted a measurement study of commonly deployed serverless functions and learned that the function performance and resource utilization depend crucially on \emph{function semantics and inputs}.
Our key insight is to delay making resource allocation decisions until after the function inputs are available.
We introduce \system, a resource management framework for serverless systems that makes decisions \emph{as late as possible} to right-size each invocation to meet functions' performance objectives (SLOs) and improve resource utilization.
\system\   uses an online learning agent to right-size each function invocation based on the features of the function input and makes cold-start-aware scheduling decisions. 
For a range of serverless functions and inputs, {\system} reduces SLO violations by 11-73\% while not wasting any vCPUs and reducing wasted memory by 64-94\% in the median case, compared to state-of-the-art systems, including Aquatope, Parrotfish, and Cypress.
\end{abstract}

\vspace{-2mm}
\section{Introduction}
\label{sec:introduction}
\vspace{-1mm}

One of the key benefits of serverless computing for developers is that it requires no management. 
Developers do not need to hand-specify their resource needs: the provider makes these decisions for them and developers can focus on their application logic alone. 
However, today's serverless systems lack performance guarantees for function invocations and end up adversely impacting developers and providers. 
For users with performance-critical applications, such as detection of videos with indecent contents uploaded to YouTube, 
unknown resource management policies that they cannot control are a problem ~\cite{BV_ServerlessComputing, Cold-Start-Overheads}. 
Meanwhile, providers lack visibility into user functions, limiting their ability to make cost-performance trade-offs on behalf of the users while optimizing resource efficiency.

The point at which resource allocation decisions are made in a function's life-cycle divides existing work into two categories: \emph{early} and \emph{late} decision-making. 
The \emph{early decision-making} systems make resource allocation decisions \emph{as soon as} a function is registered.  
Commercial serverless frameworks (AWS Lambda~\cite{AWSLAMBDA}, Azure Functions~\cite{AzureFunctions}, Google Cloud Functions~\cite{GoogleCloudFunctions}), open-source platforms (OpenWhisk~\cite{ApacheOpenWhisk}), and recent research (Parrotfish~\cite{ParrotFish}) fall under this category. 
All invocations of a function receive the same container size.
Making resource allocation decisions \emph{early} enables serverless systems to plan for the future; for example, the system can maintain a pool of pre-warm containers to mitigate cold start overheads \cite{AWS-Provisioned-Concurrency}. However, both the developer and the providers suffer due to the policies of early decision-making systems:

\begin{figure}
     \centering
     \begin{subfigure}[b]{0.2\textwidth}
         \centering
         \includegraphics[scale=0.25]{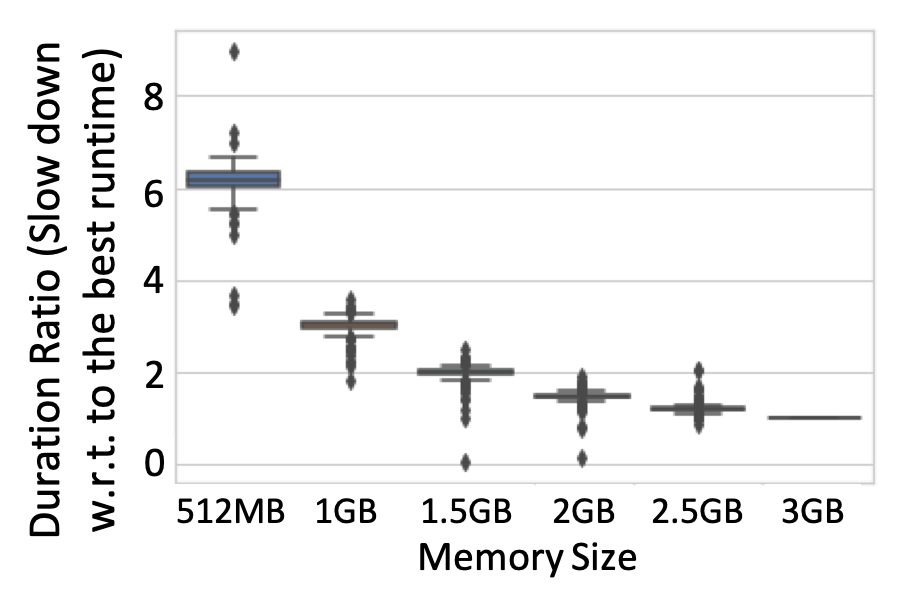}
         \caption{Slowdown w.r.t. the best runtime across mem sizes for 100 invocations of a video transcoding function.}
         \label{fig:memory_size_duration_modified}
     \end{subfigure}
     \hspace{5mm}
     \begin{subfigure}[b]{0.2\textwidth}
         \centering
         \includegraphics[scale=0.25]{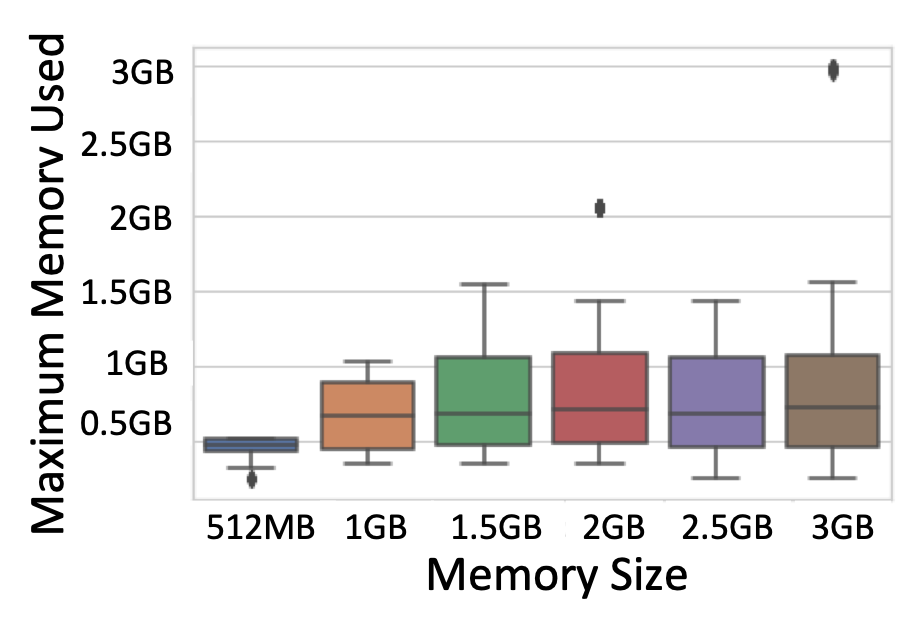}
         \caption{Maximum mem utilized vs. allocated across 100 runs of the video transcoding function (from Figure~\ref{fig:memory_size_duration_modified}).}
         \label{fig:memSize_maxMemoryUsed_new_modified}
     \end{subfigure}
     \vspace{-2mm}
        \caption{Characterizing functions with respect to the resources allocated, utilized, and performance observed.}
        \label{fig:three graphs}
        \vspace{-4mm}
\end{figure}

\noindent\textbf{1. Early and input-agnostic allocation decisions.}
Function inputs are mostly not available until invocation time. So, providers are forced to make input-agnostic resource allocation decisions.
However, across inputs, the performance and resource utilization varies greatly (see Figure~\ref{fig:three graphs}a).
Thus, premature decisions without consideration for inputs precludes optimizations, including downsizing containers for smaller inputs, which would bring significant cost benefits and improved resource utilization.
For example, if the early, function-level allocation sizes a container with 3GB of memory, and most invocations use only 1GB, the costs incurred are 2$\times$ higher for allocated but unused resources while utilization is only 33\%. 
\emph{
We argue for specialized resource allocation decisions for each input to a function.
Hence, instead of making early resource allocation decisions, we intentionally delay these decisions until after the input to an invocation is available.
}

\noindent\textbf{2. Early-binding of different resource types.}
Today's serverless platforms allocate fixed-proportions of different resource types (vCPU, memory) per function, thereby early-binding them.
However, this policy assumes that functions are both memory- and compute-intensive and leads to suboptimal allocation decisions.
For example, a compute-intensive video transcoding function wastes up to 50\% of its allocated memory (Figure~\ref{fig:three graphs}b).
\emph{Instead, we argue to make allocation decisions independently for each resource type and intentionally delay these decisions to be input-aware.}

Recently proposed \emph{late decision-making} systems, including Cypress~\cite{cypress} and Aquatope~\cite{Aquatope}, avoid making decisions at function registration to mitigate the shortcomings discussed above.
Instead of launching a container per invocation, Cypress batches multiple inputs and launches a single container for them. 
Although Cypress delays allocation decisions, it merely considers the size of the input and ignores other key input properties. We observe same-sized input videos can result in $2\times$ difference in resource utilization depending on other properties of inputs, such as video resolution (\S~\ref{sec:characterization}).
Aquatope~\cite{Aquatope} and Bilal et al.,~\cite{Bilal-Serverless} avoid early-binding of resource types and make independent decisions for vCPU and memory. However, they make input-agnostic decisions. 
Moreover, Aquatope uses a large neural network that requires offline training consuming extra resources raising the time- and resource-overheads (see \S~\ref{sec:evaluation}).



\noindent\textbf{Our Work.} We argue that the root cause of poor resource utilization and function performance is that allocation decisions for invocations are made \emph{as soon as possible} without information about the invocation needs.
Our measurement study 
shows that function semantics and input properties, not just limited to input size, greatly affect function execution time and resource utilization.

To show the importance of making delayed and independent allocations per resource type (vCPU, memory), 
we build \system, a resource management framework for serverless systems that makes resource allocation decisions \emph{as late as possible} to account for the needs of each function invocation.
{\system} introduces a performance-centric interface that allows users to specify their service-level objectives (SLOs) per invocation. 
\system\ leverages an online learning agent that predicts the required resource configuration (vCPU and memory) for each invocation to meet the user-specified SLO, using \emph{the characteristics of the function inputs}. 
\system's delayed decision-making may increase cold starts, as invocations to the same function may need differently sized containers. 
To counter this, in instances where 
a desired size is not available, \system's scheduler uses a larger container that is warm, and 
to aid future invocations, launches the perfectly sized container in the background, taking it off the critical path. 

We build {\system} atop Apache OpenWhisk~\cite{ApacheOpenWhisk}, a popular open-source serverless platform used in prior research.
We evaluate {\system} against two static baselines and three state-of-the-art systems using the Azure production trace~\cite{Serverless-In-Wild}. 
For a range of serverless functions and inputs, {\system} reduces SLO violations by 11-73\% while not wasting any vCPUs and reducing wasted memory by 64-94\% in the median case, compared to state-of-the-art systems, including Aquatope~\cite{Aquatope}, Parrotfish~\cite{ParrotFish}, and Cypress~\cite{cypress}. 

\vspace{-4mm}
\section{What Affects Function Performance and Resource Utilization?}
\label{sec:characterization}
\vspace{-3mm}




We evaluate the impact of early and late resource allocation decisions on the severity of performance variability and resource consumption of function invocations.
Specifically, we study the impact of various input properties (not limited to just input size), resource availability, and bound resource types.

We study 12 serverless functions (Table~\ref{tab:summary-functions}) from literature and benchmark suites~\cite{FunctionBench, cypress, Sebs} on top of Apache OpenWhisk (see \S~\ref{sec:evaluation} for experimental setup details).
Our functions cover a wide range of applications, from scientific workloads to data processing, and machine learning (ML) training and inference.
We collect the execution time and vCPU/memory utilization for several combinations of functions, inputs, and vCPU limits.
We run each combination 8-10 times, for a total of $\sim$8K runs across our experiments.

\vspace{-5mm}
\subsection{Impact of Function Inputs}
\label{sec: characterization-inputs}
\vspace{-2mm}

\begin{table}[bt!]
    {\footnotesize
  \centering
  \begin{tabular}{|l|c|c|c|c|}
    \hline
    \textbf{Function} & \textbf{Input Type} & \textbf{\# Runs} & \textbf{\# Sizes} & \textbf{Size Range} \\ \hline
    \emph{matmult} & square matrix & 540 & 9 & 500 - 80000 \\ \hline
    \emph{linpack} & square matrix & 660 & 11 & 500 - 8000 \\ \hline
    \emph{imageprocess} & image & 840 & 14 & 12K - 4.6M \\ \hline
    \emph{videoprocess} & video & 645 & 5 & 2.2M - 6.1M \\ \hline
    \emph{encrypt} & string & 420 & 7 &  500 - 50000 \\ \hline
    \emph{mobilenet (inf)} & image & 840 & 14 & 12K - 4.6M \\ \hline
    \emph{sentiment} & batch of strings & 716 & 12 & 50 - 3000 \\ \hline
    \emph{speech2text} & audio & 471 & 8 & 48K - 12M \\ \hline
    \emph{qr} & url & 660 & 11 & 25 - 480 \\ \hline
    \emph{lrtrain} & training set & 160 & 4 & 10M - 100M \\ \hline
    \emph{compress} & file & 434 & 7 & 64M - 2G \\ \hline
    \emph{resnet-50 (inf)} & image & 574 & 9 & 184K - 4.6M \\ \hline
  \end{tabular}
  \vspace{-3mm}
  \caption{Summary of serverless functions studied.}
  \label{tab:summary-functions}
}
\end{table}

Previous work studied the impact of the size of an input on serverless function performance~\cite{cypress}.
However, this study included only a subset of the functions we used and concluded that input size impacts function performance linearly; we find that the linearity does not hold for all the functions we studied. 
Moreover, the effect of other input properties (e.g., video resolution) on both execution time and resource utilization was not studied.
To holistically understand the impact of inputs, we consider two questions:
    (a) is there a linear relationship between input size and performance across functions?
    (b) to what extent do input properties other than size affect function performance and resource utilization?


\begin{figure}[tb!]
  \centering
  \includegraphics[width=\columnwidth]{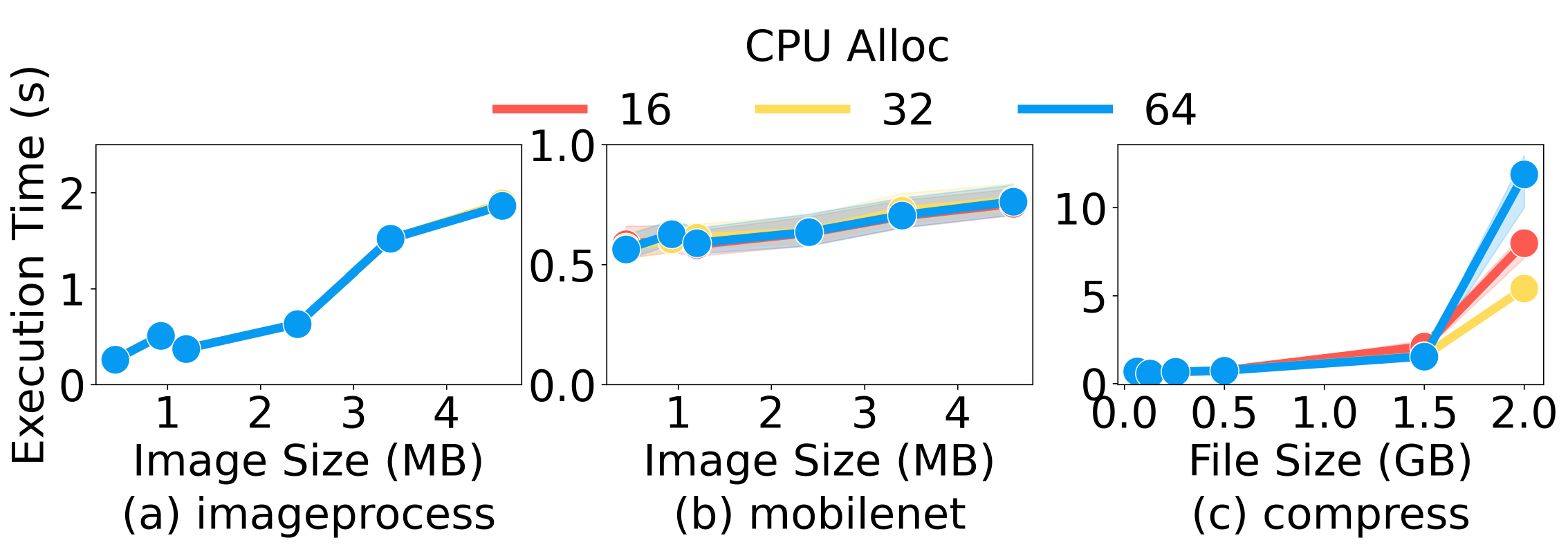}
  \vspace{-7mm}
  \caption{Input size vs. execution time showing a positive correlation for three serverless functions.}
  \label{fig:input-dependence}
\end{figure}

\noindent\textbf{Observations.} 
Figure~\ref{fig:input-dependence} presents the input size vs. execution time for three functions per vCPU allocation (we omit the other due to space constraints). 
For every function, regardless of input type, we observe that function performance is positively correlated with input size.
Moreover, larger inputs of multi-threaded functions display more variability compared to single-threaded functions; for example, with an input size of 2GB, \emph{compress} shows 50\% variability in its execution time (Figure~\ref{fig:input-dependence}c). 
However, we do not observe a consistent linear relationship between input size and execution time (\emph{imageprocess}, \emph{compress}) contrary to previous findings~\cite{cypress}.

\begin{figure}[tb!]
  \centering
  \includegraphics[width=\columnwidth]{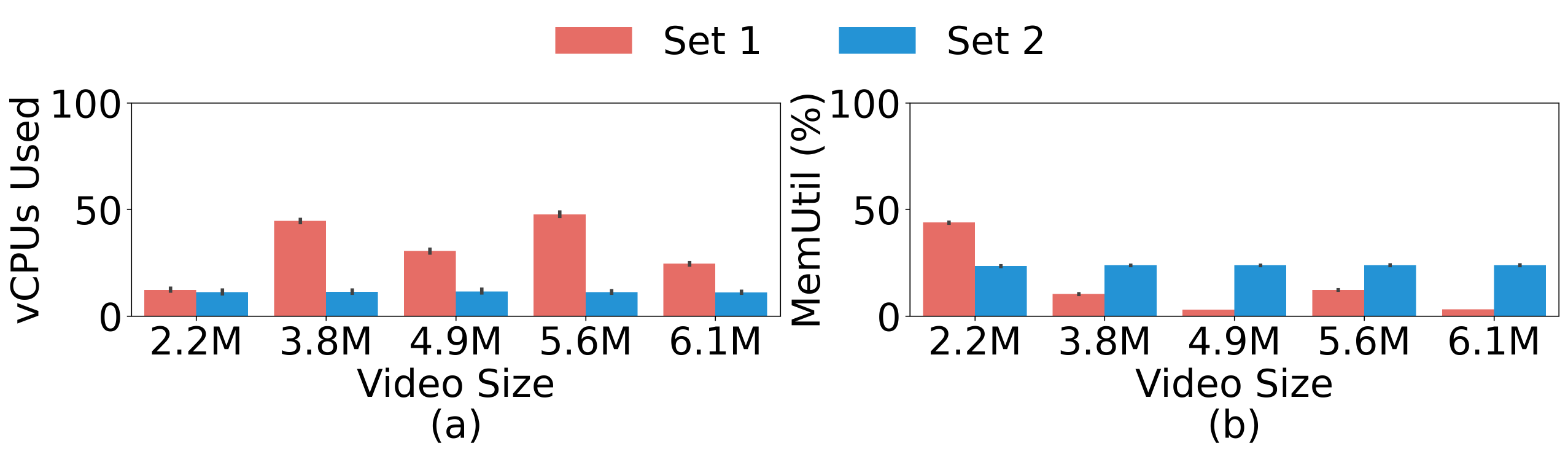}
  \vspace{-7mm}
  \caption{\emph{videoprocess}'s (a) vCPU and (b) memory utilization as a function of video size. Input properties other than size impact resource utilization (e.g., video resolution).}
  \label{fig:videoprocess}
\end{figure}

Figure~\ref{fig:videoprocess} further shows the non-linear relationship between input size and resource utilization. 
We compare the number of vCPUs used by \emph{videoprocess} on two input sets of different videos.
We see that two inputs of the same size (e.g., 3.8MB) differ by $\sim 70\%$ in the number of vCPUs used.
Moreover, while set-1 exhibits unpredictable relationships between input size and vCPUs used, set-2 exhibits constant utilization, regardless of the video size.
To understand these differences, we compare video properties beyond just size: frame rate per second, video length, bit rate, and video resolution. 
We find that resolution is the key property affecting resource utilization. 
While the resolution is constant in set-2 ($1280\times720$), it widely varies between the different video sizes in set-1. 
Inputs with higher resolutions ($1280\times720$) have lower vCPU and higher memory utilization, whereas the inverse is seen for lower-resolution inputs. 

\noindent\textbf{Takeaway \#1.} 
The effect of input properties (not just limited to size) on function performance and resource utilization is more complex than the  linear relationship previous works assume.
Hence, existing resource allocators that ignore input properties are suboptimal.
We must build an allocator that is holistically input-aware to benefit providers and developers.
\vspace{-3mm}
\subsection{Impact of Function Semantics}
\label{sec: characterization-more-resources}
\vspace{-2mm}

Previous works claim that function performance improves monotonically with additional resources~\cite{Orion}, irrespective of the function semantics.
We evaluate the extent to which this is true by observing the effect of additional resources on function performance.

\begin{figure}[tb!]
  \centering
  \includegraphics[width=\columnwidth]{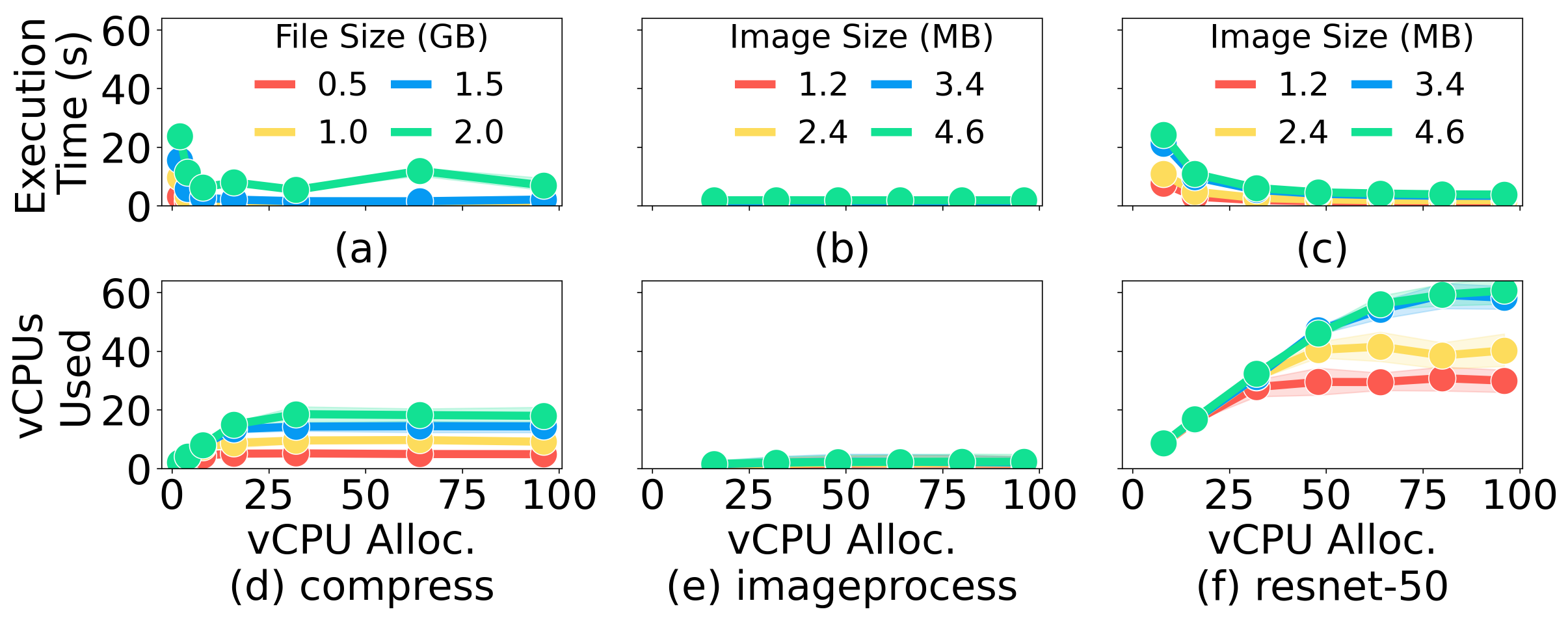}
  \vspace{-6mm}
  \caption{Execution time (top row) and vCPU utilization (bottom row) vs. vCPU allocation. Serverless functions exhibit a range of bounded parallelism.}
  \label{fig:cpu-alloc-exec}
\end{figure}

\noindent\textbf{Observations.}
\emph{compress} and \emph{resnet-50} can benefit from additional vCPUs, as the execution time decreases (Figure~\ref{fig:cpu-alloc-exec}a,~\ref{fig:cpu-alloc-exec}b).
We note that \emph{matmult, lrtrain}, and \emph{linpack} also exhibit these trends. 
However, \emph{compress} and \emph{resnet-50} show that the gains of increasing vCPU allocations saturate: Figures~\ref{fig:cpu-alloc-exec}a and~\ref{fig:cpu-alloc-exec}c exhibit plateaued vCPU utilization for a given input size.
Similar to \S ~\ref{sec: characterization-inputs}, though, we see the impact of input size on resource utilization.
Large file and image sizes exhibit higher vCPU utilization for \emph{compress} and \emph{resnet-50}, respectively.

Meanwhile, \emph{imageprocess} does not benefit from more vCPUs (Figure~\ref{fig:cpu-alloc-exec}b), even though its performance is input-dependent, as explained in \S~\ref{sec: characterization-inputs}. 
Regardless of the allocation and input size, utilization is always hovering around 1 vCPU (Figure~\ref{fig:cpu-alloc-exec}e). In fact, several of our functions exhibit this single-threaded nature: \emph{sentiment, encrypt,} and \emph{speech2text}. 
Finally, we note that while \emph{imageprocess} and \emph{resnet-50} both have the same images as their inputs, the two functions exhibit different performance and resource utilization. 

\noindent\textbf{Takeaway \#2.} Serverless platforms host single- and multi-threaded functions with potentially bounded parallelism.
Additional resources may not always help.
Hence, resource allocators should tailor their policies to suit the type of function.

\vspace{-3mm}
\subsection{Impact of Binding Resource Types}
\label{sec: characterization-coupled}
\vspace{-1mm}

We now evaluate how tightly binding different resource types impacts function performance and resource utilization.

\noindent\textbf{Observations.} Figure~\ref{fig:videoprocess} shows that \emph{videoprocess} uses up to 48 vCPUs (3.8MB input in set-1), but its memory utilization is at most 41\% (0.8GB). 
Thus, \emph{videoprocess} (and also \emph{matmult, linpack} and \emph{lrtrain}) is compute-intensive.
Meanwhile, we found \emph{sentiment} to be memory-bound, as it used 100\% of its memory but only 1 vCPU.
Cloud providers can experience severe underutilization due to tight binding early on at the time of function registration.
For example, providing enough memory to \emph{sentiment} on AWS Lambda would lead to 50\% underutilization of vCPUs. 
Meanwhile, to allocate 50 vCPUs to \emph{videoprocess} would require an $\sim85$GB memory allocation, resulting in 93\% memory underutilization.

\noindent\textbf{Takeaway \#3.} 
Early decision-making systems have little information on an invocation's resource needs.
Hence, these systems bind resource types to generalize their 
policies to a variety of functions.
Delaying resource allocation decisions until inputs are available obviates the need to bind resource types, as policies can intelligently use inputs and function semantics to make independent allocations per resource type. 

\vspace{-3mm}
\section{{\system} Design}
\label{sec:design}
\vspace{-2mm}


\begin{figure}[tb!]
  \centering
  \includegraphics[width=\columnwidth]{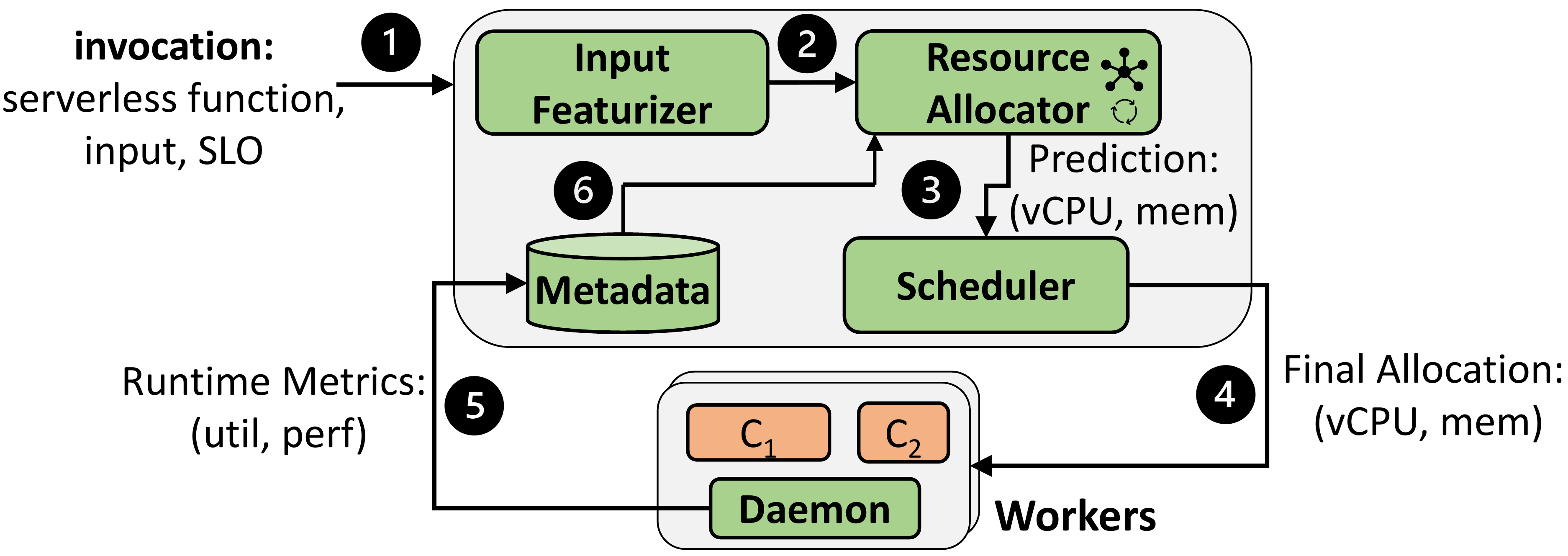}
  \vspace{-7mm}
  \caption{System architecture highlighting the life cycle of a serverless invocation through {\system}. 
  }
  \label{fig:system-architecture}
\end{figure}


We use the takeaways from \S~\ref{sec:characterization} to guide our design of {\system}, a resource management framework that makes dynamic resource allocations for each function invocation \emph{as late as possible}.
{\system}'s objective is to schedule each function invocation with the required amount of resources to meet an \emph{invocation's} SLO (execution time) while reducing the number of allocated idle vCPUs and memory.
{\system} achieves this aim through two main components:
\vspace{-2mm}
\renewcommand\labelenumi{(\theenumi)}
\begin{enumerate}
    \setlength{\itemsep}{0pt}
    \setlength{\parskip}{0pt}
    \item \textbf{Resource Allocator:} The Resource Allocator predicts the resource allocation (vCPU, memory) required to meet an invocation's SLO. \S ~\ref{sec:characterization} shows the importance of delaying resource allocation decisions until a function's input is available, but we also see the complexity in accounting for inputs and function semantics to right-size invocations. To handle this complexity, \system's Resource Allocator employs online learning to make its resource allocation predictions. We detail \system's Resource Allocator in \S ~\ref{sec:lra}.
        
    
    \item \textbf{Scheduler:} 
    Delayed resource allocations per invocation may introduce more cold starts, as invocations to the same function may need differently sized containers.
    Hence, the goal of \system's Scheduler is to mitigate cold starts introduced by making resource allocation decisions per invocation. 
    It prioritizes routing invocations to perfectly sized containers based on the predictions from \system's Resource Allocator. 
    In instances where a desired size is not available, {\system}’s
Scheduler uses a larger container that is warm, and to aid future invocations, launches the perfectly sized containers in
the background, taking it off the critical path.
We detail {\system}'s Scheduler in \S ~\ref{sec:ls}.
    
\end{enumerate}
\vspace{-2mm}


Figure~\ref{fig:system-architecture} describes {\system}'s workflow. 
\textbf{1.} With {\system}'s new interface, an invocation specifies the serverless function, its input(s), and \emph{an SLO (execution time)}.
\textbf{2.} Upon an invocation, the \emph{Input Featurizer} extracts features specific to the function's input type (e.g., image, video, matrix) if the input has not been seen before. The features of the input and the SLO are passed to \system's Resource Allocator.
\textbf{3.} The \emph{Resource Allocator} uses these input-level features to predict the required resource allocation to meet the invocation's specified SLO. 
\textbf{4.} \emph{{\system's} Scheduler} uses this prediction to make the final decision on the allocation to give the invocation. It prioritizes routing the invocation to a warm container of perfect or larger size than that specified by the prediction to avoid cold starts.
\textbf{5.} On each worker, {\system} deploys a lightweight \emph{daemon} that collects performance and resource utilization metrics per invocation. When an invocation completes, it transmits this data to a \emph{metadata store}, which {\system's} Resource Allocator uses to update its model as necessary, thereby closing the feedback loop. 
In the next two sections, we describe the core system components in more detail.

\vspace{-3mm}
\section{{\system's} Resource Allocator}
\label{sec:lra}
\vspace{-2mm}

{\system's} Resource Allocator predicts the  resources (vCPUs, memory) required to meet an invocation's SLO.
It extracts features specific to the invocation's input type (e.g., image, audio) 
and employs online machine learning (ML) to make independent allocation predictions for each resource type.



\vspace{-3mm}
\subsection{Why Use Online Machine Learning}
\label{sec: ra-online-ml}
\vspace{-2mm}


\S ~\ref{sec:characterization} shows the complexity of making accurate resource allocations. 
While input size and execution time are correlated, other input properties (e.g., image, video resolution) also impact resource utilization and performance.
Moreover, two functions with the same input can exhibit drastically different performance and utilization depending on function semantics.
Hence, we opt to use machine learning to make resource allocation predictions per invocation in a data-driven manner.

We further design \system's Resource Allocator to use \emph{online} ML for four reasons.
(1) Online ML enables our agent to observe and adapt to the dynamically changing runtime environments.
(2) Representative function inputs may not be available offline to train accurate models.
(3) If the distribution of inputs changes over time, functions themselves evolve, or the SLO requirements change, the models that are trained offline become susceptible to possible data drifts. 
(4) Training ML models offline may not generalize to new, previously unseen functions and inputs. 

\vspace{-3mm}
\subsection{Model \emph{Per Function} or \emph{Across Functions}?}
\label{sec: model-type}
\vspace{-1mm}

Given that {\system's} Resource Allocator needs to account for function semantics in its decisions, a natural question is whether to create a model per function or a single model across all functions.
A model per function may customize well and achieve high accuracy, but it may not be scalable to build as many models as functions given the resources and training data needed to build them. 
On the other hand, a single model across functions may generalize to unseen functions, but it requires extracting meaningful function-level features to distinguish between functions and provide good accuracy.
We empirically explored this choice between building a model per function vs across functions.

\noindent\textbf{A single model across functions. }
To build a single model across functions, we need to distinguish between functions using function-level features.
We chose to extract simple function-level features that may provide insight into function semantics: lines of code, number of function calls, libraries used, and API calls made to external entities (e.g., databases).
Although rather simple, this approach is more economical and practical in a serverless setting than using statistical formal methods, which are expensive to conduct~\cite{program-analysis}.
Furthermore, we also need a standard feature vector size, which is challenging to adhere to as the number of features depends on the type of the input. 
For example, for a CSV file as an input type, we extract the file size and number of rows and columns as features, whereas for an image, we extract the image width, image height, file size, number of channels, and dots per inch. 
We qualitatively and empirically evaluated three techniques to standardize the lengths of feature vectors across functions:

\emph{Use of Embeddings:} 
One commonly used approach is to learn embeddings to transform variable-sized feature vectors per input-type into a common dimensional space. 
However, each function requires its own custom transformation engine, called an encoder (another model), as the type and range of features are specific to each function. 
Hence, learning an encoder per function defeats the purpose of using embeddings in the first place, as we still need to build multiple models \cite{pml1Book}.

\begin{figure}[tb!]
  \centering
  \includegraphics[width=\columnwidth]{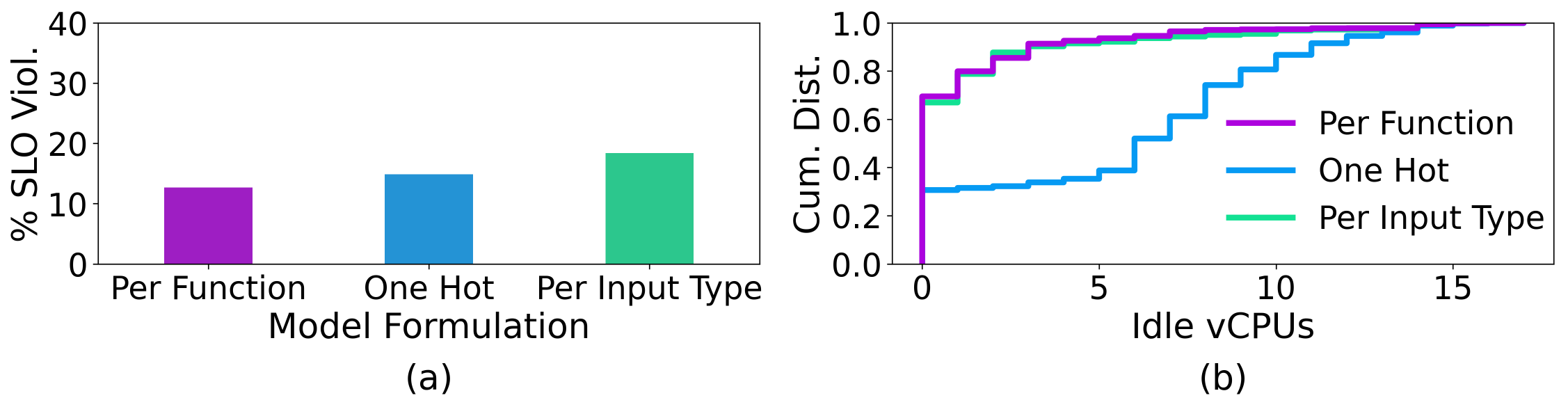}
  \vspace{-7mm}
  \caption{Analysis of three ML formulations for {\system}'s online resource allocation predictor. Building a model per function has better SLO compliance \emph{and} resource utilization than creating a model across all functions or per input type.
  }
  \label{fig:ml-formulations}
\end{figure}

\emph{One-Hot Encoding}: 
A simple and commonly used approach to standardize the length of feature vectors is to use one-hot encoding \cite{concat-and-zero}. 
We  
concatenated each function's feature vector into one large vector 
and zeroed the portion of the feature vector not belonging to the function at hand. 
We evaluated this technique against creating a model per function in Figure~\ref{fig:ml-formulations}.
Although the SLO violations are only slightly worse (Figure~\ref{fig:ml-formulations}a), its p90 idle allocated vCPUs is 5x higher than creating a model per function. 
Using one-hot encoding, the model is unable to learn for each function, so its allocation is consistently between 9-13 vCPUs.

\emph{Per Input Type}: 
Finally, we evaluated learning a model per input type (e.g., video).
With this approach, multiple serverless functions may be served by the same model.
This formulation decreases wasted vCPUs (Figure~\ref{fig:ml-formulations}b) compared to one-hot encoding but increases SLO violations (Figure~\ref{fig:ml-formulations}a). 
\emph{mobilenet} suffered from more SLO violations because this multi-threaded function consistently received one to two vCPUs.
This is because the single-threaded \emph{imageprocess} completed faster, hence the common model between the two functions first learned from the \emph{imageprocess} invocations. 
Had several \emph{mobilenet} invocations completed first, the SLO violations would decrease, but wasted vCPUs would increase, as \emph{imageprocess} allocations would surpass two vCPUs.

\noindent\textbf{One model per function. }
Rather than deploy a model that serves all functions, we evaluated building a custom model that makes resource allocation predictions for each function.
A model per function inherently accounts for function semantics and does not require any distinguishing function-level features.
Figure~\ref{fig:ml-formulations} shows the efficacy of creating \emph{one model per function}, as the model can customize to each function's needs and adapt to changes in the distribution of input data over time.
For example, even though \emph{mobilenet} and \emph{imageprocess} both have the same input type, a model per function removes the need to descriptively distinguish the semantics between two functions, enabling accurate predictions regardless of the thread-level parallelism of a function.

Heavy and complex models per function may not scale.
Moreover, offline models require collecting representative training data, which is time-consuming. 
Hence, to make it practical to create a model per function, we use an efficient, lightweight implementation of an online cost-sensitive multi-class classification algorithm.
Training a model online avoids the need to collect training data, as the model learns on-the-go from the observed performance and resource utilization per invocation.
While our model is learning, we use a large-enough default allocation. 
Moreover, we implement several safeguards to protect the system against modeling errors. 
The cost-sensitive multi-class classification algorithm requires minimal resources and makes fast predictions/updates (\S~\ref{sec:eval-overheads}).

\vspace{-3mm}
\subsection{Online Learning for Resource Allocation}
\label{sec:cpu-pred}
\vspace{-1mm}

\system's Resource Allocator uses online learning to decide how many vCPUs and memory to allocate each function invocation. 
The predicted number of vCPUs and amount of memory are then used by {\system}’s Scheduler to finalize the allocation and worker assignment for each invocation. The observed performance and utilization data is collected from the worker nodes for each invocation and stored in the metadata store. Based on this data, the online learning agent used by \system's Resource Allocator updates itself continuously as function invocations finish executions. As discussed earlier, this delayed assignment and continuous learning approach is necessary for two reasons. First, the inputs become available only at the time of invocation. Second, the availability of resources on the worker nodes may vary over time. Lastly, to make independent predictions for each resource type, we train separate online learning agents for vCPU and memory.

\vspace{-3mm}
\subsubsection{Predicting Required Number of vCPUs}
\vspace{-1mm}

We now present the formulation of the Resource Allocator's online model. 
The online model predicts the minimum number of vCPUs to allocate per invocation.

\noindent\textbf{Prediction Target.} As our goal is to meet SLOs per invocation with efficient use of resources, a natural prediction target is the minimum number of vCPUs a function invocation needs to finish execution within the specified target execution time.

\noindent\textbf{Features.}
To specialize to each input of an invocation, we use input-level features to train a per-function online learning agent. 
For each input to a function, depending on the type of the input (e.g., image, video, matrix, CSV file), we extract different features. 
For example, for images we extract the image's file size and resolution, whereas for a matrix we extract the number of rows, columns, and its density. 
See Appendix~\ref{sec:appendix-a} for the full list of the features we use per input type. 
While typical ML techniques using these input types (e.g. images, video, text) attempt to characterize the content (i.e., identify people in a video), our use case is different: our models learn the descriptive features of inputs that may affect performance and resource utilization.
We combine all this data to construct a vector for model predictions and updates.

{\system} avoids extracting input features on the critical (invocation) path whenever possible.
Any data object as an input resides in the serverless platform's datastore (e.g., AWS S3, Azure Blob Storage).
Whenever data objects are persisted in the datastore, {\system} extracts the features for that object as a background task.
The feature extraction only occurs on the critical path when a storage trigger initiates an invocation.
If an invocation is triggered without a data object as its input, we use the invocation's payload as the feature. 

\noindent\textbf{Feedback.} 
We continuously update our models by providing them feedback on the accuracy of their predictions.
To update our models, we present them the performance and resource utilization per invocation.
On each machine, we deploy a daemon that captures vCPU (and memory) utilization over an invocation's runtime.
This data, along with the execution time, is used by our cost function to update our model's weights online.

\noindent\textbf{Learning Algorithm.}
We construct the problem of predicting vCPU count as a supervised learning problem, which can be solved with either regression or classification.
In our context, regression would predict a continuous value for the minimum vCPU allocation to give an invocation. 
We note that underpredictions of this allocation are far worse than overpredictions, as underpredictions may lead to SLO violations.
Off-the-shelf implementations for regression do not naturally minimize a cost function that differentiates between underpredictions and overpredictions.
We use a \emph{cost-sensitive multi-class classification} algorithm to make our predictions, where each vCPU count represents a class.
For each class (vCPU count), we train a linear regression that predicts the cost of assigning that particular vCPU allocation to the invocation.
We select the class with the lowest cost (as per the cost function described next) as the prediction outcome.
However, this prediction is only used by the Resource Allocator if the model has observed enough invocations to the function (set by a confidence threshold hyperparameter).
If not, the Resource Allocator is not confident in the online model's predictions and defaults to a large enough number of vCPUs to allow the model to learn.

\noindent\textbf{Cost Function.} 
Our cost function updates \system's online model by leveraging the observed performance and utilization of each completed invocation.
It assigns a unique cost to each class (vCPU count)
to inform the online model of suitable vCPU allocations for the invocation.
The lowest cost it assigns is one, and 
the costs of the remaining classes grow linearly, with underpredictions being penalized further (compared to overpredictions). 
We now describe how our cost function determines the class with the least cost.

\noindent(1)\emph{ When SLO is met:} If the invocation's SLO is met, the cost function determines if fewer vCPUs than the amount allocated could also meet the SLO.
It computes a slack as the difference between the invocation's execution time and the SLO to make this decision: larger slack suggests that fewer vCPUs could still meet the SLO.
Depending on the amount of slack, the cost function either assigns the current class (vCPU allocation) or a lower class the minimum cost of one.

\noindent(2)\emph{ When SLO is not met:} 
If the invocation experienced an SLO violation, the cost function evaluates if more vCPUs are required to meet the SLO.
If the invocation used less than 90\% of its allocated vCPUs, it is unlikely that an incorrect vCPU allocation caused the SLO violation; this suggests there were other factors (e.g., variability in the system, infeasible SLO, network bandwidth contention) that caused the violation.
In this case, we assign the lowest cost to the vCPU class that was actually used by the invocation.
However, if the invocation had a high vCPU utilization, this indicates that the invocation likely requires more vCPUs to meet its SLO.
Hence, the cost function assigns the lowest cost to a class greater than the maximum vCPUs utilized. Similar to case (1), this class is determined based on the slack (or lack thereof). 

\begin{figure}[tb!]
  \centering
  \includegraphics[width=\columnwidth]{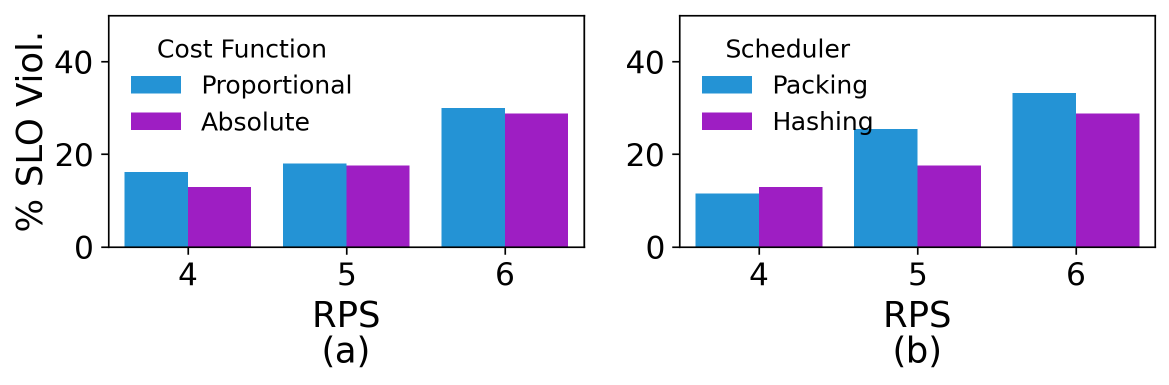}
  \vspace{-8mm}
  \caption{Design exploration of {\system}'s (a) cost function and (b) scheduler algorithm. {\system} incurs fewer SLO violations using the absolute cost function in its Resource Allocator and a hashing-based scheme in its Scheduler.
  }
  \label{fig:design-exploration}
\end{figure}

In both cases, the slack determines which class we assign the lowest cost to. 
We considered two different techniques for how to use the slack. (1) \emph{Proportional}: use the slack in proportion to the execution time of an invocation to increase or decrease the vCPU allocation, and (2) \emph{Absolute}: use the slack's absolute value to increase or decrease the vCPU allocation; for every $X$ seconds past the execution time, increase the vCPU allocation by 1 and for every $Y$ seconds below the execution time, decrease the vCPU allocation by 1. 
Figure~\ref{fig:design-exploration}a shows the percentage of invocations with SLO violations between these two techniques.
We tuned $X$ to be 0.5 seconds and $Y$ to be 1.5 seconds.
\emph{Absolute} incurs fewer SLO violations, as it is more aggressive in increasing vCPU allocations in response to SLO violations compared to \emph{Proportional}. 
This allows the online model to quickly learn by observing the efficacy of different allocations in meeting SLOs. 
We note that the p95 vCPU waste per invocation is 1 core higher with \emph{Absolute}, but  it is a modest tradeoff for reducing SLO violations by $\sim25\%$.

\vspace{-2mm}
\subsubsection{Predicting Required Amount of Memory}
\label{sec:mem-pred}
\vspace{-1mm}

Similar to vCPU predictions, the Resource Allocator uses cost-sensitive multi-class classification to predict the minimum amount of memory an invocation requires. 
Each class corresponds to an additional 128MB of memory allocation. 
However, unlike vCPU allocation, memory allocation does not affect the performance of an invocation, as serverless platforms, including AWS Lambda~\cite{AWSLAMBDA} and Azure Functions~\cite{AzureFunctions}, do not provide swap space; simply enough memory must be allocated to ensure an invocation is not abruptly killed by the \emph{Out-Of-Memory} (OOM) killer daemon on the Linux host. 
Hence, we do not include the SLO as a feature for memory predictions.
Similar to CPU predictions, we penalize underpredictions more heavily than overpredictions.
However, the cost function is much simpler, as the online model does not need to explore higher or lower memory limits in response to observed execution times. 
Instead, it assigns the lowest cost to the class corresponding to the observed memory utilization. 

To mitigate the chances of OOM exceptions, we employ two safeguards. (1) We set the memory confidence threshold to 2$\times$ the vCPU confidence threshold; the online model must observe twice the number of invocations per function before its memory predictions are used.
Increasing this confidence threshold enables our online models to learn from a variety of inputs, as the execution of both smaller and larger inputs complete within the threshold.
With this threshold, less than 1\% of invocations are killed from OOM exceptions 
(\S~\ref{sec:eval-sensitivity}).
(2) We ensure that the predicted allocation is larger than the size of the data object input(s) to the serverless function; if not, we default the memory allocation to the largest amount.

\vspace{-3mm}
\section{{\system}'s Scheduler}
\label{sec:ls}
\vspace{-2mm}


We design a new scheduler to complement {\system}'s Resource Allocator for three main reasons:
(1) {\system}'s input-aware resource allocation per invocation is likely to increase cold starts, as there may be multiple different container sizes required per function invocation to meet SLOs.
(2) Previous serverless schedulers optimize how to schedule or pack invocations to containers~\cite{Atoll, cypress, golgi, owl}, how many containers to create~\cite{golgi, cypress}, and when to create/destroy them~\cite{Aquatope, cypress}. 
Other than Hermod~\cite{Hermod}, previous schedulers do not optimize container-to-server placement, as they do not consider the exact resource usage per invocation when determining the server to place the container on. We empirically show why Hermod's policies do not work in our case in a few paragraphs.
Finally, Kubernetes-based schedulers~\cite{vhive, openfaas, knative, kubernetes} consider utilization, but do not vertically scale on a per-invocation basis like {\system}.
(3) The OpenWhisk scheduler is memory-centric: it only considers the aggregate allocated memory of invocations on a given server when load balancing.
This scheduling policy falls short when making independent resource allocation decisions per resource type; we observed that this leads to severe oversubscription of vCPUs.

We design {\system}'s Scheduler that (1) mitigates the cold starts introduced by \system's Resource Allocator by judiciously routing invocations to larger-than-optimal warm containers and (2) tracks the vCPU and memory load per server to make intelligent load balancing decisions leveraging the accurate predictions made by {\system}'s Resource Allocator.
We note that these optimizations are not specific to OpenWhisk.
{\system}'s Resource Allocator is compatible with any scheduler that makes similar optimizations.

\noindent\textbf{{\system}'s Scheduler Algorithm.}
The Scheduler mitigates cold starts by (1) prioritizing routing the invocation to a warm container of the exact size predicted by \system's online learning agent, (2) routing the invocation to a warm container larger but closest to the size predicted, and in the worst case (3) creating a new cold container of the exact size if no warm containers are suitable.
If the Scheduler routes to a larger container, it also proactively launches a container of the requested size in the background; this reduces the frequency of underutilized resources in the future.
When routing an invocation to a container, whether it is cold or warm, the Scheduler ensures the server has enough available resources. To avoid interference across functions sharing a server, we limit the overall utilization of each server following relevant literature~\cite{Borg, alicloud}. 

Upon a cold start, the Scheduler needs to pick a server to create the new container on.
Similar to OpenWhisk~\cite{FaasRank}, {\system}'s Scheduler attempts to reduce cache contention and improve locality by assigning a server to each function as its "home server".
A hashing algorithm determines the home server of each function~\cite{ApacheOpenWhisk}.
Then, {\system}'s Scheduler first attempts to create new containers on this server, however if the home server does not have capacity, it then finds the next server that does.
If no server has capacity, the Scheduler randomly chooses a server to create the container on.

OpenWhisk's hashing algorithm disperses invocations based on the function across servers.
Rather than using a hashing algorithm to determine which server to create a container on, we also implemented a policy of packing invocations onto one server until the server's capacity is reached, as proposed by Hermod~\cite{Hermod}, before routing invocations to a new server.
Figure~\ref{fig:design-exploration}b shows the SLO violations incurred between these two policies for high loads. 
At higher loads, Hermod's packing algorithm created more SLO violations.

Several of our functions, including \emph{matmult}, \emph{lrtrain}, \emph{imageprocess},  
retrieve inputs from an external database, requiring network bandwidth. 
Packing invocations per server until the server reached peak vCPU capacity made network bandwidth the bottleneck of the server, thereby degrading invocation performance.
Hermod's evaluation used functions requiring no network bandwidth, thereby showing its efficacy.
Hence, {\system} uses the hashing-based algorithm in its final design.

\noindent\textbf{Creating Idle Containers in the Background.}
Creating containers in the background introduces more idle warm containers to provision.
Several systems~\cite{cypress, golgi} use bin-packing solutions to decrease the number of containers provisioned to improve resource utilization.
We argue that the container number is the wrong resource utilization metric, as containers are simply an abstraction over the actual hardware resources.
In fact, we found that creating idle containers in the background introduces little overhead to the system because while idle, containers do not consume vCPU or memory.
Thus, by launching idle containers, the Scheduler does not increase the current load of a server. 
In {\system}, we use the default OpenWhisk keep-alive policy for these idle containers but our scheduler is compatible with more advanced approaches such as the ones proposed by Shahrad et al.~\cite{Serverless-In-Wild}.

\vspace{-3mm}
\section{{\system} Implementation}
\label{sec:implementation}
\vspace{-3mm}

We implement {\system} on Apache OpenWhisk~\cite{ApacheOpenWhisk}. 

\noindent\textbf{Implementing \system's Resource Allocator.}
We implement {\system}'s Resource Allocator as a shim layer on top of OpenWhisk in Python.
Any system can easily use {\system} by issuing gRPC requests.
The Resource Allocator interacts with OpenWhisk by issuing API calls to invoke functions.
We implement the Resource Allocator's online agent using Vowpal Wabbit~\cite{VW}, a library with an efficient online implementation of the cost-sensitive multi-class classification algorithm. 
If an invocation's input resides in a datastore (e.g., AWS S3, Azure Blob Storage), we obtain input-level features from {\system}'s in-memory metadata store 
residing on the same node.
Otherwise, we use the invocation's payload as the features. 
To allow the agent to learn an invocation's resource requirements given its inputs, we use default allocations for initial invocations (10 for vCPU predictions, 20 for memory predictions).


\noindent\textbf{Implementing Decoupled Resource Allocations.}
Similar to commercial cloud platforms~\cite{AWSLAMBDA, GoogleCloudFunctions}, OpenWhisk binds resource types; it allocates a proportional vCPU share based on the user-specified memory.
We decouple these resource types throughout OpenWhisk and make vCPU allocations a hard limit to enable delayed, independent allocations per resource type. 
We introduce a new resource limit class, \verb|CPULimit()|, that can be set for each invocation and update the OpenWhisk API to include this limit as a header in the API call.


\noindent\textbf{Implementing \system's Scheduler.}
We implement {\system}'s Scheduler in OpenWhisk in Scala.
To load balance decoupled resources, {\system}'s Scheduler considers both the aggregate vCPU and memory allocation of active invocations to determine a worker's load.
We introduce a new hyperparameter, \emph{userCPU}, to set the vCPU oversubscription limit per worker.
To track warm containers and their sizes, the Scheduler maintains a map per worker. 
The memory overhead in the worst case is two entries for each invocation if a perfectly sized container is never found. 
However, the Scheduler's proactive policy and OpenWhisk's container keep-alive policy decrease the chances of maintaining new containers over time. Hence, these maps only amount to a few megabytes. 

\noindent\textbf{Implementing \system's Daemon.}
{\system} launches a lightweight daemon with two threads on each worker. One thread collects utilization metrics per container every 10 ms using Linux cgroups and stores these metrics in an in-memory datastore.
The other thread (1) tracks completed invocations and collects their performance metrics (execution time, cold start latency) from the worker log and (2) uses gRPC to send this data together with the corresponding container utilization metrics to the Resource Allocator to update its agent online.
\vspace{-3mm}
\section{Evaluation}
\label{sec:evaluation}
\vspace{-3mm}

We evaluate {\system} by answering the following questions:
\vspace{-2mm}
\renewcommand\labelenumi{(\theenumi)}
\begin{enumerate}
    \setlength{\itemsep}{0pt}
    \setlength{\parskip}{0pt}
    \item How well does {\system} improve performance and resource utilization against baselines? ($\S$~\ref{sec:eval-e2e})
    \item How does {\system}'s Resource Allocator respond to SLO violations and resource contention? ($\S~\ref{sec:eval-ra}$)
    \item How does {\system}'s Scheduler reduce cold starts? ($\S ~\ref{sec: eval-scheduler}$)
    \item Is {\system} robust to changes in (a) the Scheduler's vCPU oversubscription limit, (b) the Allocator's confidence threshold, and (c) an invocation's SLO? ($\S ~\ref{sec:eval-sensitivity}$)
    \item What overheads does {\system} introduce? ($\S ~\ref{sec:eval-overheads}$)
\end{enumerate}

\vspace{-6mm}
\subsection{Methodology}
\label{sec: eval-methodology}
\vspace{-2mm}

\noindent\textbf{Workloads.}
We evaluate {\system} using functions (see Table~\ref{tab:summary-functions}), covering scientific applications, data processing, ML inference, and ML training. 
These functions use a mix of variable-sized data objects with multiple descriptive features. 
Finally, the functions vary in semantics (single- vs. multi-threaded) and execution time (100s of ms to a few minutes). 
We use Azure's function invocation traces~\cite{Serverless-In-Wild} to generate scaled down inter-arrival patterns of invocations.
We randomly choose a ten minute window from the trace, and for each minute, randomly generate start times for each invocation within that minute. 
We then randomly pick a subset of the start times per minute to match the requests per second (RPS) we are targeting.
Finally, we randomly choose a function/input to run at each start time. 
We evaluate {\system} against our baselines with RPS (requests per second) values 2 through 6 which corresponds to a wide load range in our setup.
This methodology aligns with several previous works~\cite{Hermod, cypress, golgi}.

\noindent\textbf{Evaluation Metrics.}
{\system} aims to meet an invocation's SLO using the minimum amount of resources. 
Hence, we use three metrics to evaluate it: (1) the percentage of SLO violations, (2) the amount of allocated but idle resources, and (3) the resource utilization per invocation. 
Every unique function/input combination has its own SLO, a property unique to {\system}.
To set an SLO, we run the function with the corresponding input in isolation on every vCPU count from 1 to 32 and obtain the median execution time across the invocations. We set the SLO to be 1.4$\times$ the median. 
This leads to different function/input combinations requiring different resource allocations to meet their SLO.
Note that this SLO is much tighter than previous works: Cypress sets the SLO by increasing \emph{the maximum execution time} observed by 20\%~\cite{cypress}. 
We show {\system}'s efficacy on a range of SLOs in \S~\ref{sec:eval-sensitivity}.



\noindent\textbf{Baselines.}
We compare {\system} to two static baselines and three state-of-the-art systems, summarized below: 
\vspace{-2mm}
\renewcommand\labelenumi{(\theenumi)}
\begin{enumerate}
    \setlength{\itemsep}{0pt}
    \setlength{\parskip}{0pt}
    \item \emph{{Static-\{Medium, Large\}~\cite{AWSLAMBDA, ApacheOpenWhisk}}}: Users often request simple static allocations, choosing a medium (12 vCPUs, 3GB mem) or large (20 vCPUs, 5GB mem) allocation per function. These baselines use OpenWhisk's default resource management and scheduling policies. 
    \item \emph{Static Allocation Developer Tools (Parrotfish~\cite{ParrotFish})}:
    Parrotfish is a state-of-the-art developer tool that assists developers in determining the optimal allocation prior to function deployment.
    We provide Parrotfish with two representative inputs (medium and large) per function and use its recommendation for all invocations. 
    
    \item \emph{Bayesian Optimization (Aquatope~\cite{Aquatope})}: 
    Bayesian optimization (BO) is a popular technique used to improve resource allocations in the cloud  \cite{CherryPick, Aquatope, Bilal-Serverless}. 
    Aquatope constructs noise and uncertainty-aware Bayesian neural networks \emph{per function} to make resource allocation decisions \emph{per function}~\cite{Aquatope}. Aquatope targets serverless workflows, but we use it for single-stage workflows  for a fair comparison. Similar to Parrotfish, we supply Aquatope with the same two representative inputs to obtain predicted allocations and request the predicted allocation for all invocations of that function. To compare resource allocators fairly, we use {\system}'s Scheduler with Aquatope's resource allocations; Aquatope decouples resource types and hence the scheduler should consider subscription of the server vCPUs. 
    \item \emph{Bin-Packing Solutions (Cypress~\cite{cypress})}: 
    Cypress makes input-size aware resource allocations by leveraging linear regression to predict an invocation's execution time and uses its prediction to assign a batch size per invocation. It then packs similarly-sized batches into the same container to minimize container provisioning. 
    We implement Cypress on top of OpenWhisk.
\end{enumerate}

\vspace{-2mm}
\noindent\textbf{Testbed.}
We use a cluster of 17 machines connected via a NetXtreme-E 10Gb/25Gb RDMA Ethernet Controller.
Every machine has two Intel Xeon Gold 6240R CPUs operating at 2.40 GHz running Ubuntu LTS 20.0.4 with the 5.4.0 Linux kernel and 192GB of memory. 
Each CPU has 48 cores. 
We use one machine to host OpenWhisk's API gateway, Controller, CouchDB, and {\system}'s central components (allocator, scheduler, metadata store). 
The rest of the machines each host an OpenWhisk Invoker and {\system}'s lightweight daemon. 
We allocate 125GB memory and 90 vCPUs to each Invoker.
We set the vCPU limit at 90 cores to keep server utilization higher than 60\%, as is the goal of Borg~\cite{Borg}.
We deploy our lightweight daemon on each Invoker to collect resource utilization metrics per invocation. 
Overall, our entire setup contains 1,440 Invoker vCPUs, evaluating {\system} and the baselines at a larger scale than previous works~\cite{Aquatope, cypress, Bilal-Serverless, ParrotFish, Hermod}.

\vspace{-3mm}
\subsection{End-to-End Results}
\label{sec:eval-e2e}
\vspace{-2mm}



\begin{figure}[tb!]
  \includegraphics[width=\columnwidth]{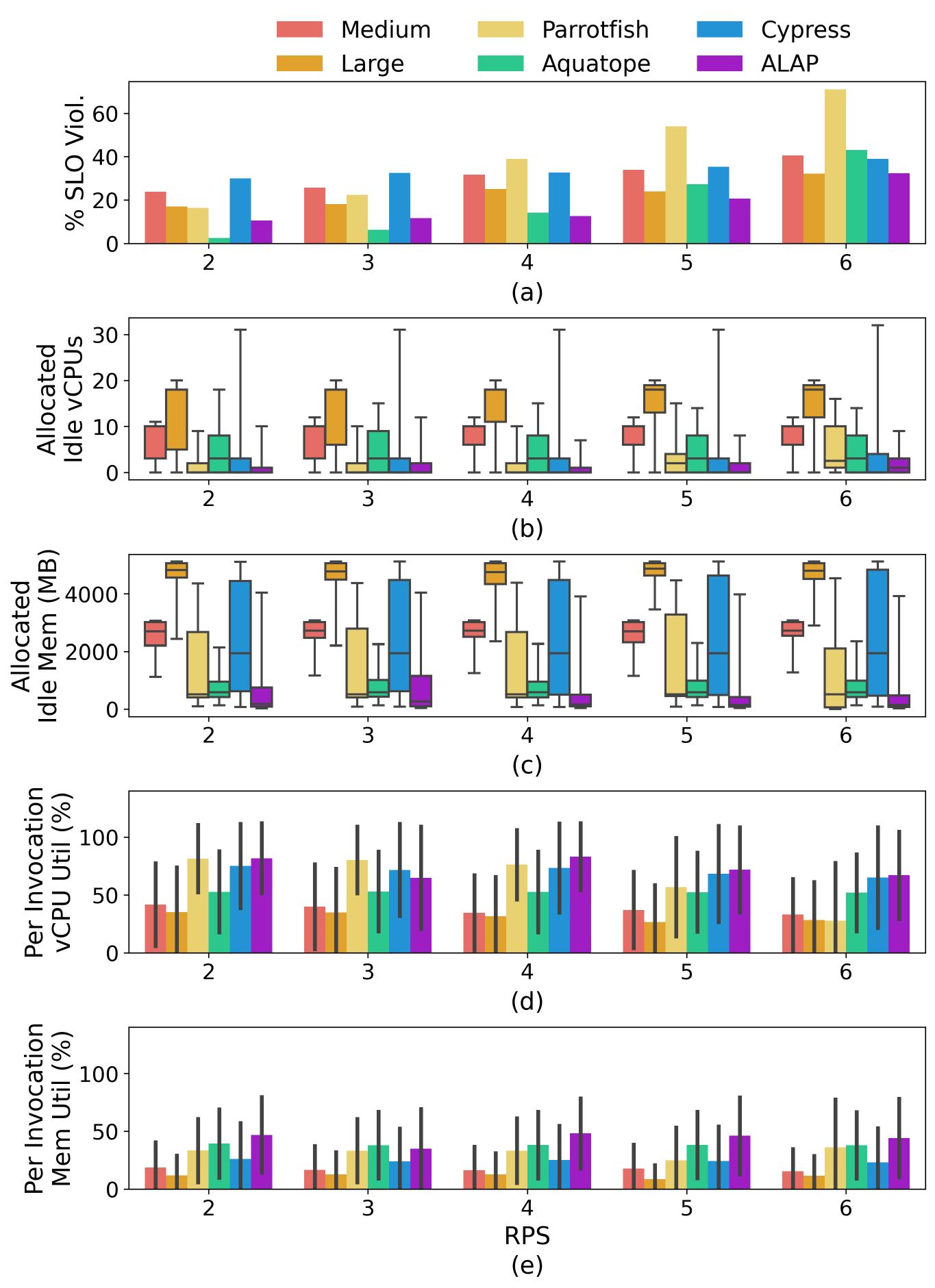}
  \vspace{-7mm}
  \caption{E2E comparison of (a) \% SLO violations, (b) wasted vCPUs per invocation, (c) wasted memory per invocation, (d) CPU utilization per invocation, and (e) memory utilization per invocation. {\system} reduces SLO violations and wasted resources while maintaining (slightly improving) resource utilization across loads. 
  }
  \label{fig:system-e2e}
\end{figure}

We begin with an end-to-end comparison of {\system} against all the baselines using our constructed traces and three evaluation metrics:
the percentage of invocations with SLO violations (Figure~\ref{fig:system-e2e}a), the wasted resources per invocation (Figures~\ref{fig:system-e2e}b,~\ref{fig:system-e2e}c), and the resource utilization per invocation (Figures~\ref{fig:system-e2e}d,~\ref{fig:system-e2e}e). 
Overall, \emph{{\system} consistently reduces SLO violations compared to the baselines while reducing wasted resources and improving resource utilization}.



\noindent\textbf{Static Baseline Analysis.} 
The static baselines show that while increasing allocated resources can improve SLO compliance (Figure~\ref{fig:system-e2e}a), invocations often do not need and hence waste larger allocations (Figures~\ref{fig:system-e2e}b,~\ref{fig:system-e2e}c). 
Although it may seem that static-large should meet every invocation's SLO, the default OpenWhisk scheduler poorly load balances due to its memory-centric policy; thus, a few servers are heavily oversubscribed with most of the invocations, causing vCPU contention and performance degradation regardless of the allocation size.

\noindent\textbf{Parrotfish Analysis.} 
Parrotfish's objective is to reduce the cost for the developer.
Parrotfish ends up allocating more memory than required to obtain more vCPUs (bound resource types) and reduce the invocation execution time.
This results in invocations consistently wasting large amounts of memory; {\system} reduces median wasted memory by $\sim4\times$ compared to Parrotfish across loads (Figure~\ref{fig:system-e2e}c).

Moreover, Parrotfish shows poor SLO compliance, as its SLO violations are 1.7$\times$ higher than static large at RPS 6 (Figure~\ref{fig:system-e2e}a).
Parrotfish's relatively tighter allocations enable OpenWhisk's scheduler to pack more invocations than static-large onto a single server, increasing vCPU contention and degrading invocation performance.
While Parrotfish achieves similar vCPU utilization to {\system} at low loads, {\system} reduces the p95 vCPU waste by $\sim1.9\times$ at RPS 6 (Figure~\ref{fig:system-e2e}b).
Thus, {\system} demonstrates the importance of making independent allocation decisions per resource type and load balancing with consideration for both vCPU and memory utilization.




\noindent\textbf{Cypress Analysis.} 
Cypress (1) assumes functions are single-threaded, (2) allocates resources proportional to an invocation's batch size, and (3) assumes frequent arrival of similarly sized invocations in order to pack them within a container.
Hence, Cypress allocates few vCPUs to every invocation (including multi-threaded ones), thereby increasing SLO violations (Figure~\ref{fig:system-e2e}a). 
Moreover, its memory utilization is poor (Figure~\ref{fig:system-e2e}c,~\ref{fig:system-e2e}e), as one invocation is allocated multiples of its memory needs under sparse arrival patterns. 

\noindent\textbf{Aquatope Analysis.} 
Aquatope ignores inputs but decouples resource types using BO. 
It suffers from fewer SLO violations at lower loads (RPS 2-3) than {\system} because the allocations lead to underutilized resources and therefore reduced interference (Figure~\ref{fig:system-e2e}a).
Aquatope wastes 3x vCPUs in the p95 at low loads compared to {\system} (Figure~\ref{fig:system-e2e}b).
{\system} learns that allocating more vCPUs to single-threaded invocations is wasteful. 
With high loads (RPS 5-6), {\system} reduces SLO violations by 25\% compared to Aquatope, while maintaining good resource utilization due to tighter allocations (Figures~\ref{fig:system-e2e}a-c).
Though several inputs can utilize more vCPUs if allocated, larger allocations are not required to meet SLOs.
Aquatope's larger resource allocations enable invocations that can meet SLOs with fewer resources to "steal" resources from invocations that require them to meet SLOs.
Meanwhile, {\system} learns to provide just enough resources to meet an SLO, reducing vCPU contention and improving SLO compliance.

Finally, we note that {\system}'s p95 wasted memory is higher than Aquatope (Figure~\ref{fig:system-e2e}c). 
To ensure invocations do not suffer from OOM exceptions, {\system} requires its online agent to learn from 20 invocations per function before its memory prediction can be used (\S~\ref{sec:eval-sensitivity}). 
Hence, 20 invocations per function are allocated the default maximum amount of memory (4GB).
After these 20 invocations, {\system} makes accurate predictions and minimizes wasted memory, as {\system} reduces the median and p75 idle memory by 4x and 2x, respectively, compared to Aquatope at high loads (Figure~\ref{fig:system-e2e}c).

\begin{figure}[tb!]
  \includegraphics[width=\columnwidth]{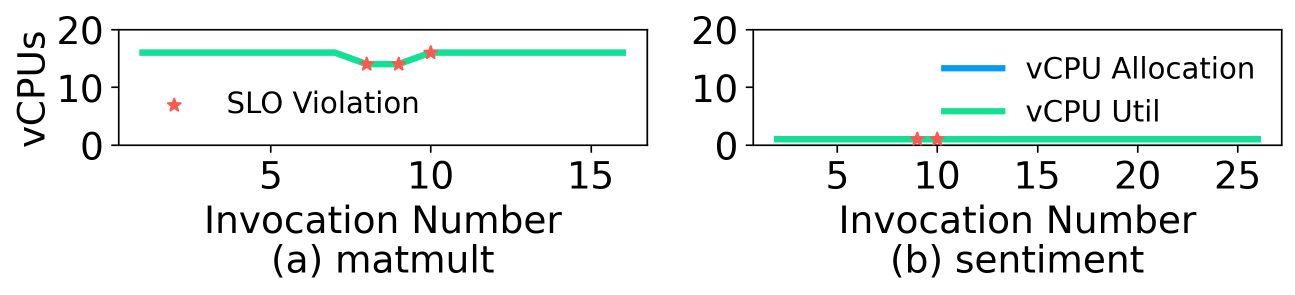}
  \vspace{-7mm}
  \caption{Zoomed-in timeline showing the allocated and utilized cores of one input for (a) \emph{matmult} (multi-threaded) and (b) \emph{sentiment} (single-threaded). {\system} responds to SLO violations by increasing vCPU allocations only if the function has enough parallelism to exploit extra resources.
  }
  \label{fig:zoomed-one-input}
\end{figure}
\vspace{-3mm}
\subsection{{\system}'s Response to SLO Violations}
\label{sec:eval-ra}
\vspace{-2mm}

We analyze {\system}'s allocation predictions for different functions and in response to SLO violations.
Table~\ref{tab:no-containers} (Appendix~\ref{sec:appendix-b}) shows the number of unique container configurations allocated per function.
This number is consistently low across RPS values for single-threaded functions (e.g., \emph{imageprocess}, \emph{sentiment}).
Most variation comes from different memory requirements per input.
Meanwhile, the number of unique sizes grows for multi-threaded functions with the RPS value (e.g., 45 for \emph{matmult} with RPS 6).
We explain this with Figure~\ref{fig:zoomed-one-input}, which shows the vCPU allocations, utilizations, and SLO violations of one input for \emph{matmult} (Figure~\ref{fig:zoomed-one-input}a) and \emph{sentiment} (Figure~\ref{fig:zoomed-one-input}b) over time.
{\system} attempts to lower the vCPU allocation for \emph{matmult}'s input on the seventh invocation to explore whether a smaller invocation can meet the SLO.
However, after observing SLO violations, it quickly reverts back to 15 vCPUs to meet the SLO.
Meanwhile, {\system} does not increase the allocation for \emph{sentiment} upon an SLO violation, as it learns the function is single-threaded and cannot use more vCPUs. 
Hence, the Resource Allocator creates more unique container sizes to explore multiple allocations for multi-threaded functions, while stabilizing the allocation for single-threaded functions.

\vspace{-4mm}
\subsection{\system's Scheduler Reduces Cold Starts}
\label{sec: eval-scheduler}
\vspace{-2mm}


\begin{figure}[tb!]
  \includegraphics[width=\columnwidth]{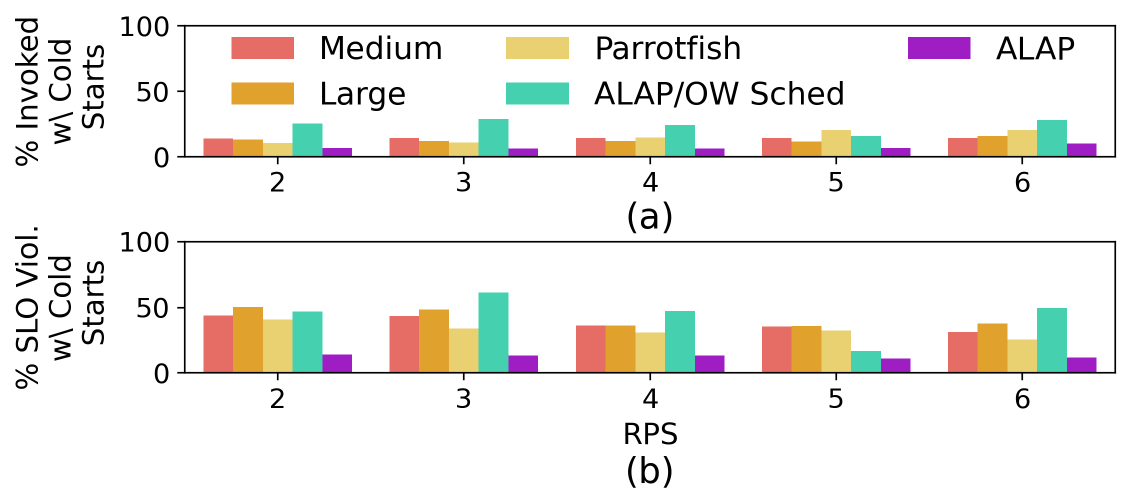}
  \vspace{-8mm}
  \caption{Comparison of (a) \% of invocations with cold starts and (b) \% of SLO violations that had cold starts. {\system}'s Scheduler makes delayed, independent allocations practical. 
  }
  \label{fig:cs-comparison}
\end{figure}

We now focus on \system's Scheduler, and show its contribution to mitigating cold starts and meeting SLOs.
We compare the percentage of invocations (Figure~\ref{fig:cs-comparison}a) and SLO violations (Figure~\ref{fig:cs-comparison}b) with cold starts between {\system}, \system's Resource Allocator with the default OpenWhisk scheduler, and the baselines.
We omit Cypress and Aquatope as we do not implement their warm container pool policies. 
Overall, regardless of the load, \system's Scheduler is instrumental in meeting SLO violations: it reduces the percentage of invocations with cold starts by 50\% compared to {\system} with the default OpenWhisk scheduler and by 56\% compared to Parrotfish (RPS 6).
In fact, most SLO violations using {\system} are not due to cold starts (Figure~\ref{fig:cs-comparison}b), but rather because of vCPU contention and variability.
Thus, making delayed resource allocations by itself is not enough to meet SLOs while improving resource utilization; \system's Scheduler mitigates the potential cold starts introduced with allocations that are specialized to function inputs.

\vspace{-3mm}
\subsection{Sensitivity Analysis of {\system}}
\label{sec:eval-sensitivity}
\vspace{-2mm}

\begin{figure}[tb!]
  \includegraphics[width=\columnwidth]{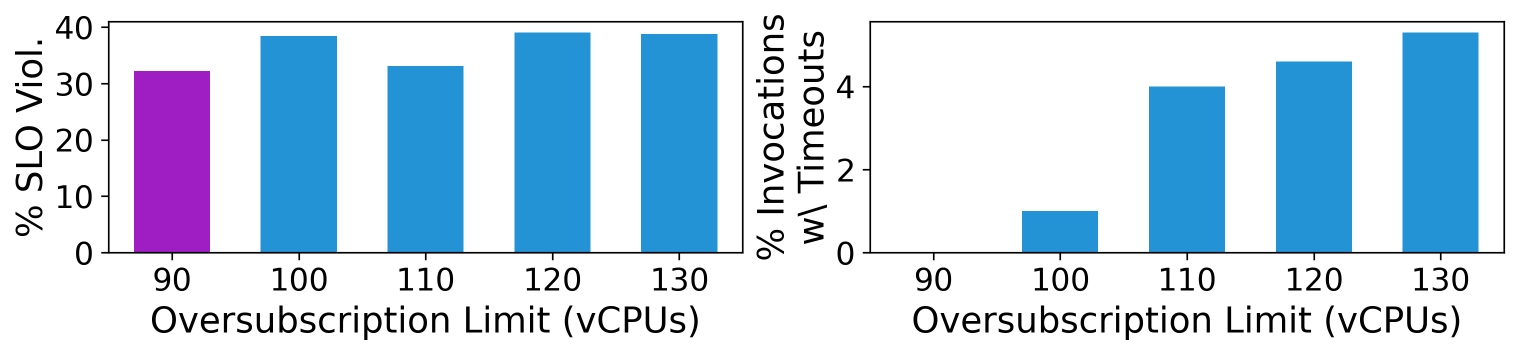}
  \vspace{-7mm}
  \caption{{\system}'s sensitivity to variation in the vCPU oversubscription limit per worker. RPS is set to 6. 
  }
  \label{fig:sensitivity-cpu}
\end{figure}

\system\  uses two key hyperparameters: (1) vCPU oversubscription limit and (2) prediction confidence. 
We evaluate \system's sensitivity to these parameters across a range of SLOs. 

\noindent\textbf{vCPU Oversubscription Limit.} 
The oversubscription limit sets the maximum allowable capacity on each server. 
Figure~\ref{fig:sensitivity-cpu}a shows that increasing this limit of a worker above 90 vCPUs
does not decrease SLO violations (at RPS 6).
Note that our servers have 96 vCPUs.
In fact, as we increase the oversubscription limit, more invocations timeout, where no response is sent back to the user (5.3\% of invocations timeout with a oversubscription limit of 130).
With higher vCPU oversubscription limits, {\system}'s Scheduler packs more invocations onto fewer servers.
Hence, compute resources are heavily contended and several invocations time out.
This shows the simplicity of setting this hyperparameter to achieve high utilization and SLO compliance: set it close to the number of cores on the server.

\begin{figure}[tb!]
  \includegraphics[width=\columnwidth]{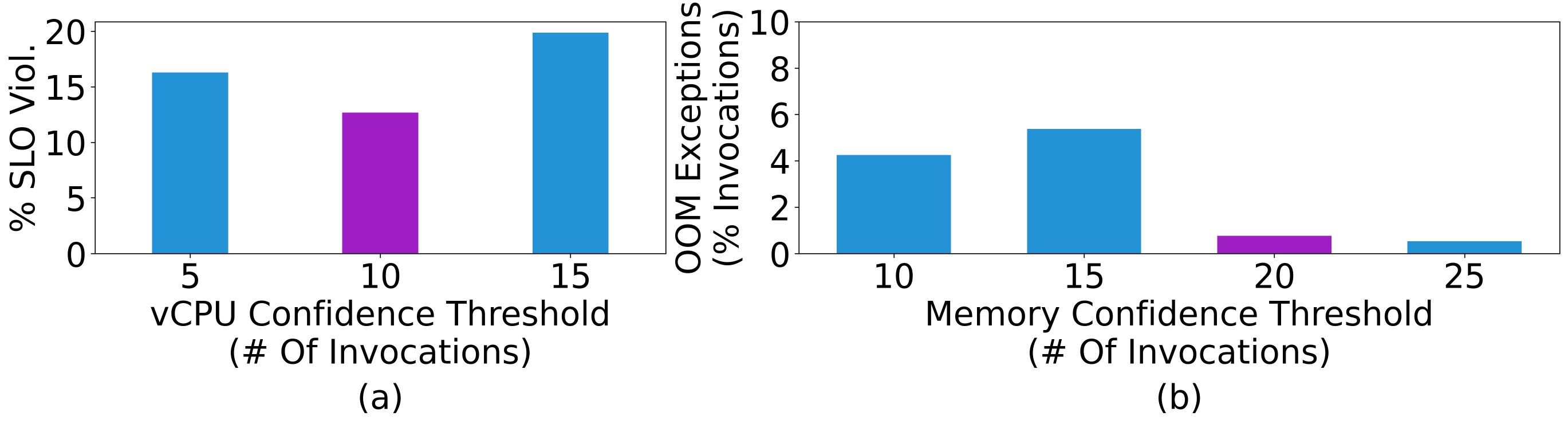}
  \vspace{-7mm}
  \caption{{\system}'s sensitivity to changes in the confidence threshold for (a) vCPU and (b) memory prediction. Purple is the threshold we set. 
  }
  \label{fig:sensitivity-confidence}
\end{figure}

\noindent\textbf{Confidence Threshold.} 
The confidence threshold is the required number of invocations the online model must observe as it trains before the Resource Allocator uses its predictions.
We show SLO violations for various vCPU confidence thresholds (Figure~\ref{fig:sensitivity-confidence}a) and the percent of invocations killed due to OOM exceptions with different memory confidence thresholds (Figure~\ref{fig:sensitivity-confidence}b).
Lower is better for both plots.
Increasing the memory confidence threshold significantly reduces the percentage of invocations killed ($<1\%$ at 20 invocations and above).
Hence, continuously increasing the memory threshold will reduce OOM exceptions further.
However, increasing the vCPU confidence threshold does not necessarily help reduce SLO violations.
More invocations use containers with the default 16 vCPUs when increasing the confidence threshold.
Hence, similar to Aquatope, multi-threaded invocations "steal" resources from smaller invocations, creating increased vCPU interference and SLO violations.
Although the exact length of the learning phase depends on the complexity of the function and inputs, setting the vCPU confidence threshold to 8-12 invocations was sufficient for all cases we observed.



\begin{figure}[tb!]
  \includegraphics[width=\columnwidth]{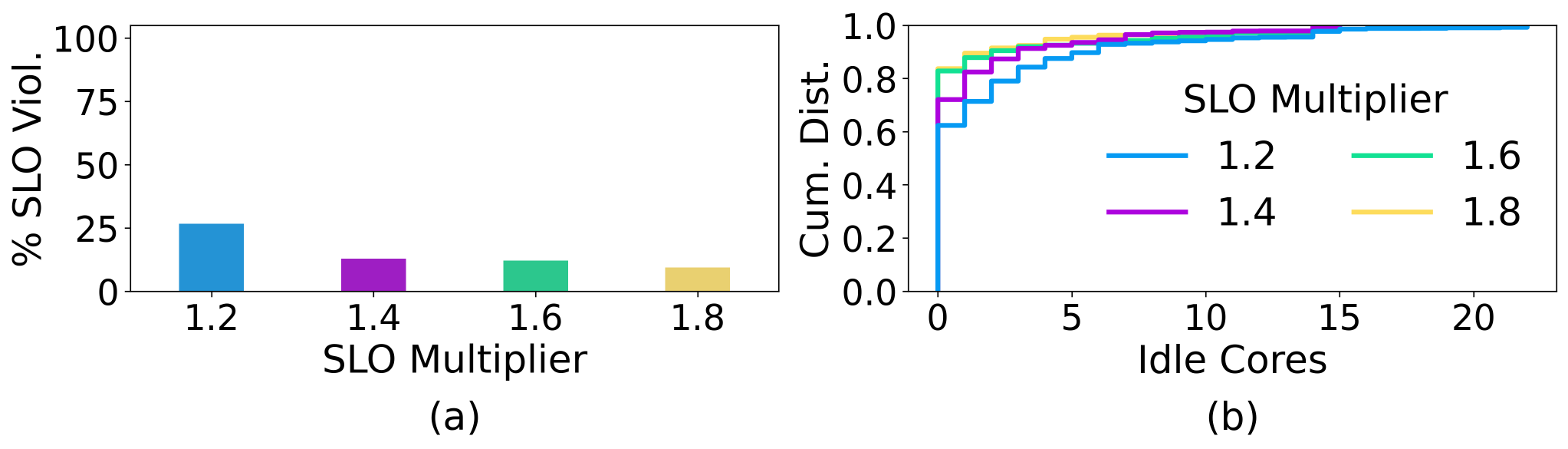}
  \vspace{-7mm}
  \caption{Analysis of (a) \% SLO violations and (b) the distribution of idle vCPUs produced by {\system} for different SLO multipliers. Our evaluation sets the SLO to 1.4$\times$.}
  \label{fig:sensitivity-slo}
\end{figure}

\noindent\textbf{Varying SLOs.}
We evaluate {\system}'s ability to handle stricter and relaxed SLOs.
We set the SLO multiplier to 1.4$\times$ throughout the rest of the evaluation.
Stricter SLOs (smaller multiplier) result in more SLO violations (Figure~\ref{fig:sensitivity-slo}a).
To meet strict SLOs, {\system} increases allocation sizes for functions capable of using more vCPUs (multi-threaded functions).
However, for single-threaded functions, {\system} learns to not unnecessarily allocate more vCPUs, as meeting the SLO is simply infeasible in a shared environment.
The median idle vCPUs does not increase, and the p95 increases by 2 vCPUs with the strictest SLO (1.2$\times$) compared to the most relaxed SLO (1.8$\times$). 
Hence, {\system} is robust across a range of SLO values. 
\vspace{-3mm}
    \subsection{Overheads}
\label{sec:eval-overheads}
\vspace{-2mm}

\begin{figure}[tb!]
  \includegraphics[width=\columnwidth]{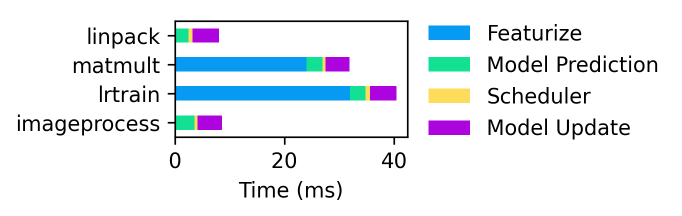}
  \vspace*{-9mm}
  \caption{Overheads introduced by {\system}.}
  \label{fig:overheads}
\end{figure}

We examine the overheads introduced by {\system} (Figure~\ref{fig:overheads}).
Overall, regardless of the function and input type, model prediction (2-4 ms) and {\system}'s Scheduler (0.5-1.5 ms) contribute little to the overall overheads.
Model updates take 4-5 ms, however, updates are not on an invocation's critical path.

{\system}'s overhead mostly comes from input featurization. 
Featurizing \emph{matmult} and \emph{lrtrain} inputs takes 20-35 ms, 
whereas it only takes 0.13 ms to featurize \emph{imageprocess}'s inputs.
\emph{linpack} does not require any featurization, as the {\system} agent's feature vector comprises the payload to the invocation.
\emph{matmult} and \emph{lrtrain} have higher overheads as the Featurizer needs to open input files to obtain features (row size, col size),  whereas {\system} can obtain metadata about images (resolution, no. of channels) without opening its files.
However, as mentioned in $\S$~\ref{sec:lra}, the overheads of featurization are only on the critical path if an object store triggers an invocation.
Moreover, many inputs (video, image, audio) will not have large featurization overheads as obtaining metadata about the files does not require file operations. 
Hence, we believe the improvements in resource utilization and SLO compliance outweigh the small overheads introduced by delayed allocations.

\vspace{-3mm}
\section{Existing Approaches}
\label{sec:background}
\vspace{-3mm}

Several systems address the inefficiencies mentioned in \S \ref{sec:characterization}, but none offer a solution covering all three problems.

\noindent\textbf{Offline Profiling Tools.}
Tools, such as AWS Lambda Power Tuning~\cite{AWSLambdaPowerTuning} and Parrotfish~\cite{ParrotFish}, assist developers in determining the optimal memory limit to specify upon function creation, reducing the profiling costs to users.
But, such offline decisions typically lead to underutilization as they must pessimistically allocate enough resources for all input sizes.
Moreover, such tools are time-consuming (several minutes): it took $\sim$25 minutes to profile one function with Parrotfish.

\noindent\textbf{Bin-Packing Solutions.}
Several systems over-commit container instances by packing multiple invocations.
Cypress~\cite{cypress} batches multiple invocations with similar slacks together. However, Cypress still ignores inputs, as the base allocation is determined per function. 
Moreover, such batching approaches suffer from poor resource utilization under sparse arrival patterns of real-world serverless workloads~\cite{Serverless-In-Wild}. 
Golgi~\cite{golgi} maintains separate pools of non-overcommitted and overcomitted instances.
Its scheduler predicts the execution time of incoming invocations to determine which instance type to route an invocation to.
Although Golgi decreases container provisioning, the system still makes input-agnostic resource allocations and couples resource types. 
Moreover, users are still required to specify the memory allocation per function. 

\noindent\textbf{Learning-Based Solutions.}
Several systems use some form of machine learning to make resource allocation or execution time predictions.
Cypress~\cite{cypress} uses linear regression to predict the execution time using the function input's size.
However, we show in Section~\ref{sec: characterization-inputs} that properties other than input size (e.g., resolution of images or videos) significantly impact execution time and resource utilization. 
Parrotfish~\cite{ParrotFish} uses parametric regression to construct a cost model per function with offline samples.
Both Parrotfish and Cypress are susceptible to data drift because the regressors are constructed offline.
Golgi~\cite{golgi} constructs online Random Forests to predict execution time, however, its predictions are input-agnostic.
Moreover, it constructs several models per function, as the models reside on server nodes and manage at most 100 containers. 
Bilal et al.,~\cite{Bilal-Serverless} use Bayesian Optimization (BO) and Aquatope~\cite{Aquatope} uses BO neural networks to predict resource allocations.
BO is a popular technique used throughout cloud systems to tune knobs and resource configurations~\cite{CherryPick}.
However, {\system} achieves superior SLO compliance and resource utilization while using lighter-weight modeling techniques.

\noindent\textbf{Serverless Optimizations.} Several systems optimize other aspects of serverless computing: reducing cold-start overheads~\cite{firecracker,container-loading,slacker,cntr,faasnet,cfs,wharf,catalyzer,sand,sock,icebreaker,rund,faasnap,fireworks,hotstarts,lukewarm}, efficient function-to-function communication~\cite{faasflow,sonic,sand,pocket,locus,faastlane,rdma-fork,xdt}, and state management~\cite{boki,jiffy,fisc,halfmoon,lambdaobjects,apiary,mu2sls,ofc,faast}.
{\system} is compatible with these optimizations.
\vspace{-3mm}
\section{Conclusion}
\label{sec:conclusion}
\vspace{-3mm}
We presented {\system}, a resource management framework for serverless systems with a performance-centric interface, that meets user-specified SLOs for function invocations while improving resource utilization. To do so, \system\ delays resource allocation decisions until inputs are available to account for the resource needs of each input to a function. Further, \system\  makes independent decisions for each resource type. \system\  uses an online learning agent that predicts the required amount of resources to meet an invocation's SLO. \system's Scheduler mitigates cold starts by proactively launching right-sized containers and prioritizing the use of warm containers.
Across a diverse set of serverless functions, {\system} outperforms several other state-of-the-art systems in reducing SLO violations while increasing resource utilization.

\bibliographystyle{plain}
\bibliography{references}

\clearpage
\appendix
\section{Model Inputs}
\label{sec:appendix-a}

\begin{table}[tb!]
    {
\footnotesize
  \begin{tabular}{l|ccccc}
    \toprule
    \textbf{Input Type }& \textbf{Features} \\
    \midrule
    \textbf{image} & image width, image height, num. channels \\
                   & x-axis dots per inch (dpi), y-axis dpi, filesize \\[4pt]
    \textbf{matrix} & num. rows, num. columns, density \\[4pt]
    \textbf{video} &  video width, video height, duration, bitrate \\
    & average frame rate, video encoding (e.g., mp4, mpeg4, etc.) \\[6pt]
    \textbf{csv file} & num. rows, num. columns, file size\\[4pt]
    \textbf{json} &  length of outer most object, file size\\[4pt]
    \textbf{audio} & num. channels, sample rate, duration, bit rate, \\ & whether audio is compressed using FLAC \\
    \bottomrule
  \end{tabular}
  \caption{Features Extraced per Unique Input Type}
  \label{tab:input-features}
}
\end{table}

As part of {\system}'s Resource Allocator, we build a Featurizer that extracts features from the serverless inputs. 
These features are used by the online cost-sensitive multi-class classification model to predict the required amount of vCPUs and memory to allocate an invocation to meet the invocation's SLO. 
We list the features extracted per input type in Table~\ref{tab:input-features}.
For several input types, we use linux tools to extract features: ffprobe for audio and video~\cite{ffprobe} and imagemagick for images~\cite{imagemagick}.
For other input types (e.g., matrix, csv file, json), the Input Featurizer opens the file to extract its features.


\section{Unique Container Sizes per Function}
\label{sec:appendix-b}
Table~\ref{tab:no-containers} shows the number of unique container sizes created by {\system} per function.

\begin{table}[tb!]
    {
\footnotesize
  \begin{tabular}{lccccc}
    \toprule
    & \textbf{RPS 2} & \textbf{RPS 3} & \textbf{RPS 4} & \textbf{RPS 5} & \textbf{RPS 6} \\
    \midrule
    \textbf{matmult} & 23 & 22 & 37 & 38 & 45 \\
    \textbf{encrypt} & 9 & 7 & 10 & 9 & 11 \\
    \textbf{linpack} & 39 & 39 & 44 & 39 & 55 \\
    \textbf{imageprocess} & 7 & 9 & 9 & 7 & 17 \\
    \textbf{sentiment} & 5 & 5 & 5 & 8 & 10 \\
    \textbf{mobilenet} & 5 & 8 & 11 & 12 & 14 \\
    \textbf{videoprocess} & 10 & 8 & 7 & 8 & 11 \\
    \textbf{lrtrain} & 12 & 22 & 24 & 21 & 19 \\
    \bottomrule
  \end{tabular}
  \caption{No. of Unique Container Sizes Per Function}
  \label{tab:no-containers}
}
\end{table}

\end{document}